# Role of bias voltage and tunneling current in the perpendicular displacements of freestanding graphene via scanning tunneling microscopy

Peng Xu, Steven D. Barber, Matthew L. Ackerman, J. Kevin Schoelz, and Paul M. Thibado[a),b)]

Department of Physics, University of Arkansas, Fayetteville, Arkansas 72701

[a)]American Vacuum Society member.
[b)]Electronic mail: thibado@uark.edu

Systematic displacement measurements of freestanding graphene as a function of applied bias voltage and tunneling current setpoint using scanning tunneling microscopy (STM) are presented. When the bias voltage is increased the graphene approaches the STM tip, while, on the other hand, when the tunneling current is increased the graphene contracts from the STM tip. To understand the role of the bias voltage, we quantitatively model the attractive force between the tip and the sample using electrostatics. For the tunneling current, we qualitatively model the contraction of the graphene using entropic concepts. These complementary results enhance the understanding of each other and highlight peculiarities of the system.

# I. INTRODUCTION

Because graphene is extremely thin yet possesses a high intrinsic strength and large conductivity, it has been identified as an ideal candidate material for a future generation of flexible electronics.[1] Both its elastic and electronic properties are therefore of great importance. The key to successfully investigating its mechanical attributes has been the ability to probe graphene while suspended over holes and trenches. For example, Lee *et al.* were able to measure the Young's modulus and breaking point of freestanding graphene using atomic force microscopy on exfoliated graphene laid over microholes previously etched into a Si substrate.[2] Scanning tunneling microscopy (STM) experiments have also demonstrated the ability to interact with graphene nano-membranes using forces provided by an STM tip.[3, 4] In a similar study, Klimov *et al.* took scanning tunneling spectroscopy measurements on exfoliated graphene suspended over trenches in a $SiO_2$ substrate, forming graphene drumheads which were back-gated to control carrier concentration and its deflection.[5] In these studies, the graphene was found to interact strongly with the STM tip.

STM, however, has the added potential to heat the sample through its tunneling current. Graphene is known to have an exceptionally large current-carrying capacity, and is thermally limited to about 3 $\mu$A/nm before the sample breaks down.[6, 7] Graphene also possesses a negative thermal expansion coefficient ($\alpha \approx -5 \times 10^{-6}$ K$^{-1}$), so it contracts rather than expands when heated.[8-10] Fundamentally, thermal excitations increase the amplitude of atomic vibrations; however, for single-layer graphene, it naturally favors out-of-plane movement, which results in an overall in-plane contraction. Since the tunneling current is injected into a small area just under the STM tip, this should allow one to explore local heating effects on graphene, and therefore develop a better understanding of its thermodynamic properties.

In this paper, the controlled manipulation of freestanding graphene using feedback-on STM is presented. Specifically, tip height versus bias voltage and tip height versus tunneling current measurements are investigated. It is shown that increasing the bias voltage causes the graphene to stretch toward the tip, while increasing the tunneling current causes the graphene to contract away from the tip. With the feedback electronics operational, however, we can precisely track this movement, which is orders-of-magnitude larger than typical electronic effects on stationary graphene.

## II. EXPERIMENT

Our graphene sample was grown via chemical vapor deposition,[11] then transferred onto a 2000-mesh, ultrafine copper grid comprised of square holes 7.5 μm wide and bar supports 5 μm wide. This grid was mounted on a flat tantalum STM sample plate using silver paint and transferred through a load-lock into the STM chamber, where it was electrically grounded for all experiments. The STM used was an Omicron ultrahigh-vacuum (base pressure is $10^{-10}$ mbar), low-temperature model system, operated at room temperature. STM tips were manufactured in-house by electrochemical etching of polycrystalline tungsten wire using a custom double-lamella setup with automatic gravity cutoff.[12] After etching, the tips were gently rinsed with distilled water, briefly dipped in a concentrated hydrofluoric acid solution to remove surface oxides, and then immediately introduced to the vacuum system through a load-lock.[13]

Data acquired included images of freestanding graphene, as well as feedback-on tip height ($Z$) measurements as a function of bias voltage ($V$) and tunneling current setpoint ($I_0$). "Feedback-on" means that the feedback loop controlling the vertical motion of the STM tip remained operational, allowing the system to constantly seek the height needed to provide the requested tunneling current. Each height measurement was obtained at a chosen point on the freestanding part of the sample by pausing a topography scan that was already in progress. The imaging parameters were set to a scan size of only 0.1 nm × 0.1 nm and a scan rate of only 0.1 nm/s, so that the tip effectively remained at the same location throughout multiple data runs.

During $Z(V)$ measurements, the tip bias was varied, and the vertical displacements required to maintain a constant tunneling current were recorded. On a stationary surface, this process is sometimes used to probe the density of states (DOS) at a surface. However, because the sample is grounded, the tip bias also induces an image charge, resulting in an electrostatic attraction that increases with the bias. When performing $Z(V)$ measurements on a flexible surface, such as freestanding graphene, this attraction can result in large physical movement of the sample, orders of magnitude greater than the electronic DOS effects. Thus the measurements instead reflect electromechanical attributes of the system.

During $Z(I)$ measurements, the tip bias was held constant, and the current setpoint was incremented in small steps while recording the vertical displacement required to achieve each current setpoint at each step. This measurement is more typically performed to determine the work function of a stationary sample.[14] However, the tunneling current also induces Joule heating in the surface,[15] which causes freestanding graphene to contract due to its negative thermal expansion properties. Similar to the $Z(V)$ measurements, this creates a large-scale physical movement orders of magnitude larger than the work function features. Thus the measurements instead reflect thermo-mechanical attributes of the system.

## III. RESULTS AND DISCUSSION

### A. STM Data Measurements

A 6 nm × 6 nm filled-state STM image of suspended graphene, taken with a tip bias of 0.1 V at a setpoint current of 1.0 nA, is displayed in Fig. 1(a). The overall topography is characterized by a ridge running from top to bottom, and the distinctive honeycomb lattice can be seen in the center of the image. The features appear somewhat fuzzy and stretched, however, when compared to those of graphene on a substrate due to the relative instability of the freestanding graphene sample. Next, a cross-sectional schematic of the $Z(V)$ and $Z(I)$ measurements is displayed in Fig. 1(b). The graphene sample is grounded, while the STM tip is held at some positive bias relative to the ground (note, negative tip biases also result in attraction, as expected). The tunneling current is monitored by a feedback loop, which is used to maintain a constant current between the tip and sample by adjusting the vertical position of the tip as necessary. As the tip bias is increased, for example, the sample moves toward the tip due to an increased electrostatic attraction, as shown in the figure, and the tip must retract.

Constant-current $Z(V)$ results for several current setpoints ($I_0$) are displayed in Fig. 1(c). Each curve is arbitrarily offset to have a height of zero at the starting bias voltage of 0.1 V. They are generally characterized by a quick rise as the tip bias is ramped from 0.1 V to around 1.0 V, followed by a plateau region over the remaining range to 3.0 V. The final heights increase with increasing current, reaching as high as 20 nm, indicating that larger currents result in larger total displacements. The inset plots the measured tunneling current (I) versus bias voltage on a semi-log graph, confirming that tunneling was maintained throughout the experiments. It also verifies that the graphene moved with the tip throughout the measurement, since such large changes in Z would cause I to exponentially decrease to zero on a stationary surface.

Constant-voltage $Z(I)$ results for several biases setpoints ($V_0$) are displayed in Fig. 1(d). Again, each curve is arbitrarily offset to have a height of zero at the starting current of 0.01 nA. In contrast with the $Z(V)$ measurements, the tip height decreases over the scan rather than increases. Interestingly, a small $V_0$ (0.1 V) results in a very large initial drop of approximately 15 nm from 0.01 nA to 0.10 nA, followed by a slower decrease spanning more than 5 nm before reaching the final current of 1.0 nA. At high biases ($V_0 \geq 1$ V), on the other hand, the tip height decreases only a few nm or less, appearing almost constant by comparison, and the curves for intermediate voltages fall somewhere between these two extremes. The inset, showing the measured tunneling current (I) plotted versus the setpoint current ($I_0$), confirms that the tunneling current setpoint was continuous maintained throughout the scan.

## B. Electrostatic force calculations

In order to better understand the $Z(V)$ results, we calculated the electrostatic attractive force between the STM tip and the grounded sample as a function of the bias voltage using a highly idealized model. Here, the STM tip is simplified to be a conducting sphere of radius $a$, held at potential $V$, and placed with its outer surface a distance $d$ from an infinite grounded conducting plane representing the graphene sample, as illustrated in Fig. 2(a). The resulting electrostatic problem has an exact solution found using the iterative method of images technique.[16] Because a point charge generates spherical equipotentials, the boundary condition on the sphere can be easily satisfied by removing the physical surface and placing a point charge of magnitude $q_0 = 4\pi\varepsilon_0 aV$ at the former center of the sphere. Next, to satisfy the grounded boundary condition at the plane, an equal and opposite charge $q_0$ must be placed equidistant from the plane and on the other side [see Fig. 2(a)]. Now the boundary condition on the plane is satisfied, but that on the sphere is no longer satisfied. Another positive charge of lesser magnitude must be placed inside the spherical boundary at a different location to satisfy the spherical equipotential again, followed by another opposite charge across the plane. As the iterations proceed, the magnitudes of the charges decrease so that only a finite number of image pairs (e.g., 40) needs to be considered. For illustration, three such pairs are shown in Fig. 2(a).

To determine the electrostatic attractive force between the tip and sample from the equivalent method-of-images construction, one must first computes the energy stored in the system. The electrostatic energy of the $N+1$ positive charges in the upper half-plane is given by

$$U = \tfrac{1}{2}(q_0 + \sum_{i=1}^{N} q_i)V \tag{1}$$

where $q_0$ is defined above, and

$$q_i = \frac{a}{x_0 + x_{i-1}} q_{i-1} \tag{2}$$

is the magnitude of the $i^{th}$ charge in the upper half-plane; $x_0 = a + d$ is the position of the initial charge at the sphere center, and

$$x_i = x_0 - \frac{a^2}{x_0 + x_{i-1}} q_{i-1} \tag{3}$$

is the position of the $i^{th}$ charge measured relative to the plane. Because the plane is grounded, the electrostatic energy of the charges in the lower half-plane is zero, and the above expression is the energy of the entire system. The attractive force $F$ between the sphere and plane is given by

$$-\frac{\partial U}{\partial x} = -\frac{V}{2}\sum_{i=1}^{N}\frac{\partial q_i}{\partial x} = -2\pi\varepsilon_0 aV^2 \sum_{i=1}^{N}\frac{\partial s_i}{\partial x} \qquad (4)$$

where $s_i \equiv q_i/q_0$. The last step is completed in order to pull out the common $q_0$ factor, and thereby show that the force is predicted to increase quadratically with the tip bias.

The quantitative relationship between the electrostatic force and tip bias for a number of experimentally reasonable tip radii $a$, tip-sample separations $d$, and over the voltage range used in our experiments are shown in Fig. 2(b). For clarity, the $d = 0.5$ nm and $d = 1.0$ nm family of curves are offset upward by 15 nN and 5 nN, respectively. The largest forces are generated by a smaller $d$ and larger $a$, with a maximum of 25 nN. From several scanning electron microscopy images taken in numerous studies on electrochemically etched tungsten STM tips, our best estimate for a realistic tip radius is $a = 20$ nm.[17] Similarly, from first-principles STM tunneling studies[18] we conclude that a good average tip-sample separation in these experiments is $d = 1.0$ nm. These parameters give rise to a maximum force of about 5 nN.

Up to this point, our model neglects the small changes in tip-sample tunneling junction width, $d$ which are known to occur as the tip bias is varied. However, these small changes can have a noticeable impact on the electrostatic force. To correct for this, experimental tip-height, $d_{exp}$ versus bias voltage curves were acquired for graphene on copper, and are shown in Fig. 2(c). These measurements were performed in exactly the same manner as the $Z(V)$ data earlier, and in fact, the curves bear a strong resemblance. In this case, though, the graphene is bonded to the copper,[19] preventing mechanical movement of the sample, so the tip displacements are entirely electronic and therefore, they are an order of magnitude smaller than those shown earlier in Fig. 1. The initial tip-sample separation $d_0$ in these experiments is still unknown, so we define the true $d(V) = d_0 + d_{exp}$. This relationship is then assumed to hold for the freestanding graphene, so that in the above equations $x_0(V) = a + d(V)$, and at each $V$ a new set of image charges is generated. Since the $q_i/q_0$ terms now vary with $V$, the $F(V)$ relationship is no longer purely quadratic. Three corrected $F_c(V)$ curves using $a = 20$ nm and $d_0 = 0.5$ nm are shown in Fig. 2(d). These are the final set of curves used to predict the electrostatic force as a function of bias voltage for our three setpoint currents.

## C. Discussion

The $Z(V)$ data from Fig. 1(c) can now be reduced using our model. First, according to the $Z(I)$ measurements in Fig. 1(d), the sample height is approximately the same across all currents when $V_0 = 3.0$ V. As a result, the $Z(V)$ curves, which all end at 3 V, should have nearly equal final heights rather than the same starting heights. Therefore, they have been replotted in Fig. 3(a), using the $Z(I)$ data ($V_0 = 3.0$ V) shown in the inset. Note, the shift is also independently consistent with the Z(I) data for lower voltages. This representation of the data naturally shows that the graphene plateaus at nearly the same height at the maximum voltage, representing a taut graphene configuration.

Next, the corrected force versus voltage curves derived by the electrostatic model and shown in Fig. 2(d) were used to transform the experimental $Z(V)$ curves in Fig. 3(a) into force versus height curves, F(Z) as shown in Fig. 3(b). The F(Z) curves look like reflected and rotated versions of the $Z(V)$ curves. They are approximately flat up to around $Z = 20$ nm, after which the force increases very rapidly, up to a maximum (which is just over 4 nN for $I_0 = 0.01$ nA curve). This indicates that the graphene is more flexible at first and suddenly becomes rigid. Indicating that bond bending is primarily responsible for the movement and not bond stretching. Also, each curve nearly overlaps the others, which suggests that the force necessary to bend the membrane a given amount in this region is largely independent of the tunneling current chosen. However, the starting height is directly affected by the tunneling current, and it decreases with increasing $I_0$. Finally, the work, W done by the STM on the freestanding graphene for each curve can be calculated and is found to be roughly the same for each at about 50 eV.

In order to better understand the contraction of the graphene as the current is increased, we introduce a second model for the graphene membrane as shown in Fig. 3(c,d). This illustrates a common statistical mechanics problem involving the average length and tension of a rubber band at various temperatures (*T*) and loads (F).[20] The rubber band is modeled as a series of segments which can be oriented either parallel or antiparallel (details not shown in Fig.).[21] As the rubber band is heated, the segments favor an antiparallel arrangement, decreasing the total length and resulting in a negative thermal expansion coefficient. Quantitatively, the length of a rubber band carrying a constant load is proportional in first order to *F/T*, where a larger F stretches the system and a larger T contracts the system. For our experiment, the tunneling current sets the temperature of the sample through Joule heating ($T \sim I^2$), so as the setpoint current is increased the sample's local temperature is increased. We observe that the graphene contracts sharply at first and then more slowly, creating *Z(I)* curves that qualitatively obey a $1/I^2$ relationship. In this way, the rubber band model correctly mimics both the contraction of the graphene and the specific 1/T dependence. In addition, we can estimate the temperature of the graphene directly under the STM tip. Using a value of $-10^{-5}$ K$^{-1}$ for the thermal expansion coefficient, our height contraction of 20 nm, and assuming a distance of 3 μm to the copper support (which is held at room temperature), an increase of 10-100 K is estimated.[22]

The rubber band model can also be used to explain the bias dependence of the $Z(I)$ measurements shown in Fig. 1(d). For the rubber band model, under a small load, it contracts a large amount $\Delta Z$ for a given increase in $T$, as illustrated in Fig. 3(c). However, when the load is increased, the length will change much less for the same temperature difference, $\Delta T$ (and for the same amount of work done) as shown in Fig. 3(d). For our experiments, the bias voltage sets the load or force through electrostatics, so as the voltage increases the force increases. Thus, at low bias, the small force on the graphene allows a large displacement as the current is increased [see Fig. 1(d)], whereas at high bias the large load on the graphene results in a much smaller displacement over the same increase in current.

## IV. SUMMARY AND CONCLUSIONS

A series of tip-height measurements were collected using feedback-on STM at a single location on a sample of freestanding graphene by systematically changing the bias voltage and tunneling current. Increasing the bias voltage by 3 V resulted in the graphene being attracted to the STM tip and moving as much as a 20 nm. One the other hand, increasing the setpoint current by 1 nA resulted in the graphene contracting away from the STM tip and also moving as much as 20 nm, in but the opposite direction. A simple electrostatic model was used to quantify the attractive force between the tip and the sample, which was estimated to rise as high as 4 nN and explained the experimental dependence on bias voltage. The freestanding graphene was also modeled as an entropic rubber band to successfully account for its contraction as the current was increased and raised the temperature of the sample through Joule heating. The voltage and current data sets were shown to be consistent with each other and to complement the understanding of one another. These results also highlight the very different interaction between an STM tip and freestanding graphene when compared to that between a probe tip and a stationary sample. Finally, this technique allows one to locally quantify the unusual properties arising from graphene's negative thermal expansion coefficient.

## ACKNOWLEDGMENTS

This work was supported in part by the Office of Naval Research under Grant number N00014-10-1-0181 and the National Science Foundation under Grant number DMR-0855358.

**Figure Captions**

Figure 1. (a) 6 nm × 6 nm filled-state STM image of freestanding graphene, taken with a tip bias of 0.1 V at a setpoint current of 1.0 nA. (b) Cross-sectional schematic of the $Z(V)$ and $Z(I)$ measurements performed on suspended graphene using feedback-on electronics. (c) Tip height versus bias voltage curves for three different tunneling current setpoints. Inset shows the measured tunneling currents were constant. (d) Tip height versus tunneling current curves for six different constant bias voltages. Inset confirms the measured current agreed with the setpoint as it was varied.

Figure 2. (a) To-scale diagram illustrating the iterative method-of-images technique described in the text to approximate the tip-sample system and calculate the attractive force. (b) Theoretical sphere-plane forces as a function of the sphere potential for multiple sphere radii $a$ and sphere-plane separations $d$. (c) Experimental tip height versus bias voltage curves at three different current setpoints for graphene on copper (used to determine how $d$ changes with $V$). (d) Final corrected $F_c(V)$ curves, letting $d$ begin at 0.5 nm and changing with $V$, and with $a = 20$ nm.

Figure 3. (a) The same $Z(V)$ data as in Fig. 1(c), but offset to agree with the $Z(I)$ results in Fig. 1(d). The inset shows the 3.0 V curve used to set the final heights relative to one another. (b) Force exerted on graphene by the tip as a function of height, deduced from the data shown in (a). Each curve roughly follows the same path. (c,d) Schematic diagrams illustrating the entropic rubber band model, to which freestanding graphene is compared. The same change in temperature occurs in both, resulting in a larger

displacement for the smaller weight and a smaller displacement for a larger weight (for the same amount of work, W, done).

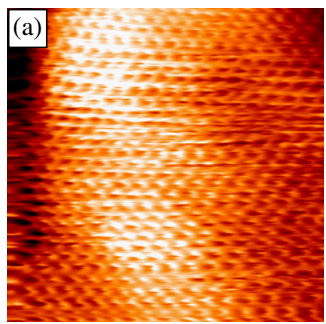
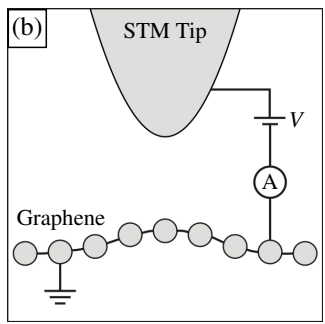
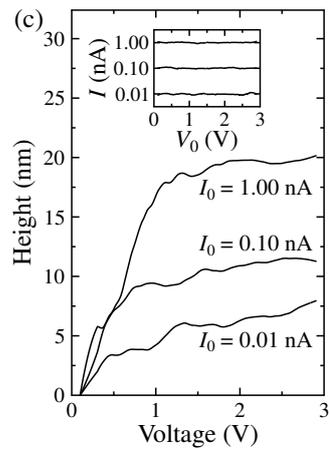
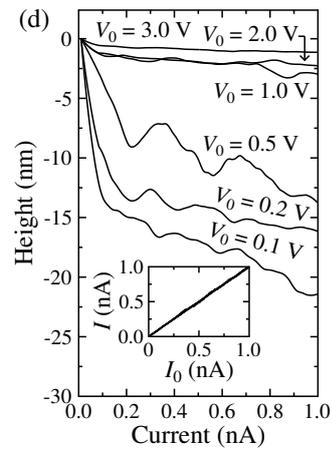

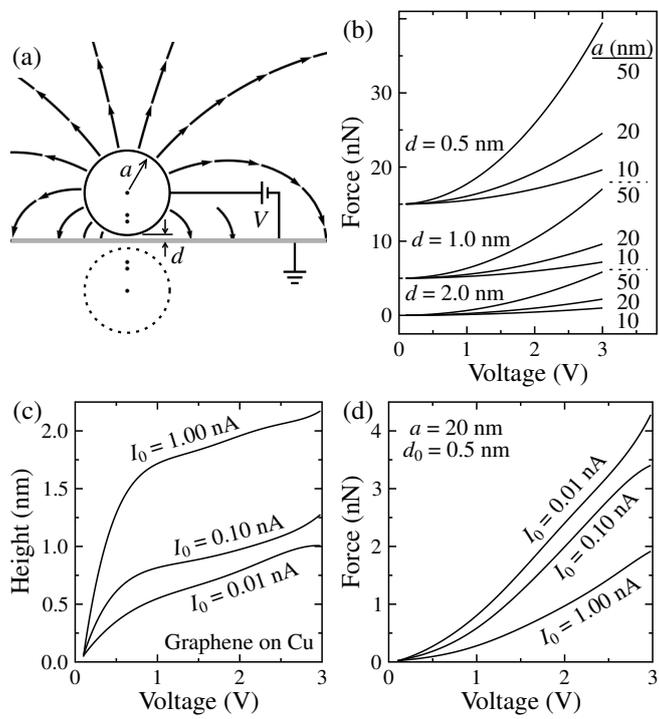

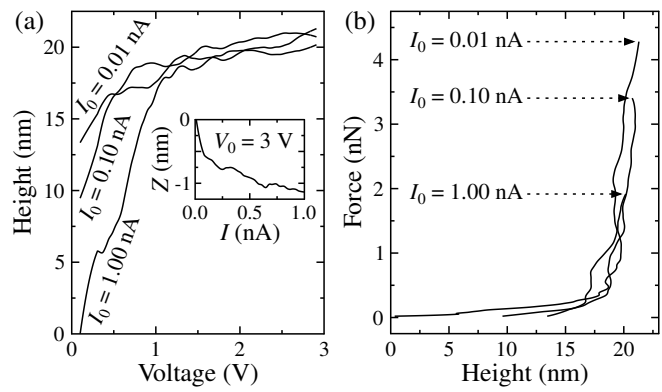
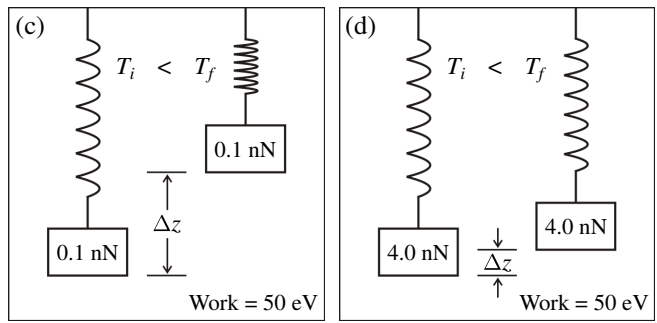